\def\year{2019}\relax
\documentclass[letterpaper]{article} % DO NOT CHANGE THIS
\usepackage{aaai19}  % DO NOT CHANGE THIS
\usepackage{times}  % DO NOT CHANGE THIS
\usepackage{helvet} % DO NOT CHANGE THIS
\usepackage{courier}  % DO NOT CHANGE THIS
\usepackage[hyphens]{url}  % DO NOT CHANGE THIS
\usepackage{graphicx} % DO NOT CHANGE THIS
\def\UrlFont{\rm}  % DO NOT CHANGE THIS
\usepackage{graphicx}  % DO NOT CHANGE THIS
\usepackage{hyperref}

\usepackage{times}  %Required
\usepackage{helvet}  %Required
\usepackage{courier}  %Required
\usepackage{url}  %Required
\usepackage{graphicx}  %Required
\usepackage{multirow}

\usepackage[sort,comma,authoryear]{natbib}

\usepackage{times}
\usepackage{helvet}
\usepackage{courier}
\usepackage{graphicx}
\usepackage{pdfpages}
\usepackage{url}
\usepackage{caption}
\usepackage{subcaption}
\usepackage{multirow}

\usepackage{amsmath}
\usepackage{amsfonts}
\usepackage{amssymb}

\usepackage{textpos}

\usepackage{xspace}
\newcommand*{\eg}{e.g.\@\xspace}
\newcommand*{\ie}{i.e.\@\xspace}

\begin{document}

%\author[2]{Author E\thanks{E.E@university.edu}}

%\title{Understanding Political Ideology of Legislators through Social Media Photographs}
%\thanks{This work was supported by the National Science Foundation under Grant No. 1831848, Hellman Fellowship, and UCLA Faculty Career Development Award}}


% % The correct dates will be entered by the editor

\title{Understanding the Political Ideology of Legislators from Social Media Images}

%\thanks{To appear in the Proceedings of International AAAI Conference on Web and Social Media (ICWSM 2020)}

\author{Nan Xi,\textsuperscript{\rm 1*} Di Ma,\textsuperscript{\rm 1*} Marcus Liou,\textsuperscript{\rm 1}\thanks{Equal contributions.} Zachary C. Steinert-Threlkeld,\textsuperscript{\rm 1} \\
%\thanks{Primarily Mike Hamilton of the Live Oak Press, LLC, with help from the AAAI Publications Committee}\\ 
\Large \textbf{Jason Anastasopoulos,\textsuperscript{\rm 2$\dagger$} and Jungseock Joo\textsuperscript{\rm 1}}\thanks{Corresponding authors.} \\ 
%\Large \textbf{Sunil Issar, J. Scott Penberthy, George Ferguson, Hans Guesgen}\\ % All authors must be in the same font size and format. Use \Large and \textbf to achieve this result when breaking a line
\textsuperscript{\rm 1}University of California, Los Angeles %If you have multiple authors and multiple affiliations
~~~\textsuperscript{\rm 2}University of Georgia\\ %If you have multiple authors % use superscripts in text and roman font to identify them. For example, Sunil Issar,\textsuperscript{\rm 2} J. Scott Penberthy\textsuperscript{\rm 3} George Ferguson,\textsuperscript{\rm 4} Hans Guesgen\textsuperscript{\rm 5}. Note that the comma should be placed BEFORE the superscript for optimum readability
% 2275 East Bayshore Road, Suite 160\\
% Palo Alto, California 94303\\
%publications19@aaai.org % email address must be in roman text type, not monospace or sans serif
jjoo@comm.ucla.edu,~~~ljanastas@uga.edu
}

\maketitle
%\vspace{-30pt}

% \begin{textblock*}{3cm}(11cm,-5.2cm)
%   \fbox{\footnotesize to appear in journal x}
% \end{textblock*}

\begin{textblock*}{25cm}(2cm,-7cm)
{\footnotesize To appear in the Proceedings of International AAAI Conference on Web and Social Media (ICWSM 2020)}
\end{textblock*}

\begin{abstract}
In this paper, we seek to understand how politicians use images to express  ideological rhetoric through Facebook images posted by members of the U.S. House and Senate.  In the era of social media,  politics has become saturated with imagery, a potent and emotionally salient form of political rhetoric which has been used by politicians and political organizations to influence public sentiment and voting behavior for well over a century. To date, however,  little is known about how images are used as political rhetoric. Using deep learning techniques to automatically predict Republican or Democratic party affiliation solely from the Facebook photographs of the members of the 114th U.S. Congress, we demonstrate that predicted class probabilities from our model function as an accurate proxy of the political ideology of images along a left--right (liberal--conservative) dimension. After controlling for the gender and race of politicians, our method achieves an accuracy of 59.28\% from single photographs and 82.35\% when aggregating scores from multiple photographs (up to 150) of the same person. To better understand image content distinguishing liberal from conservative images, we also perform in-depth content analyses of the photographs. Our findings suggest that conservatives tend to use more images supporting status quo political institutions and hierarchy maintenance, featuring individuals from dominant social groups, and displaying greater happiness than liberals. 
%\keywords{Political Ideology \and Social Media Photographs \and Legislators \and Emotion \and DW-NOMINATE }
\end{abstract}

\section{Introduction}
% Outline
% What is ideology and how does it relate to visiual communication?  
%  -  Annual review piece on ideology https://www.annualreviews.org/doi/pdf/10.1146/annurev.psych.60.110707.163600
% How might politicians use images to communicate ideological content?
% What research has been done in the past on measuring the ideological orientations of politicians and what are we adding to this literature?

% What is ideology and how does it relate to visiual communication?
% Write more here

% Up front, talk more about the connection between images, emotions and politics. 

Images generate rapid, emotional reactions~\citep{turvey1973peripheral,pessoa2002attentional}. As a result, they have served as potent tools of persuasion in a diverse array of fields from advertising and marketing to political campaigns. While images are used for diverse ends in the political sphere, they are  powerful tools of persuasion and political rhetoric because they connect emotions to politics in a way that other mediums, such as text, are unable to.  Indeed, the power of images to persuade and influence was understood well before the field of psychology even existed. At the beginning of the American Civil War, for instance, politicians used  images to elicit support for the war and to influence public sentiment on both sides~\citep{zeller2005blue}.  There is perhaps no better evidence of the importance of imagery than the observation that practically every member of the United States Congress has a \textit{Facebook} and an \textit{Instagram} account, replete with images that they share with their followers on a daily basis. 
 %A century later, during the Vietnam War, photojournalists laid bare the horrors of warfare through decimated landscapes and gruesome images of the dead and wounded in an effort to end the war.

%In the early 2000s,  the advent of social media platforms such as \textit{Twitter}, \textit{Facebook}, and \textit{Instagram} ushered in a new era of image based communication in the political sphere,  which some have dubbed ``The Age of the Image''~\citep{apkon2013age}.  

%Yet despite the importance of imagery, very little is known about how images are used as a tool of political rhetoric and persuasion. In particular, many political messages reflect the ideological stances of politicians, and political images can also visually highlight ideological values that will be appreciated by their supporters.

In this paper, we develop an automated methodology for learning about how political ideology is conveyed through images.  This methodology generates insights about the ideological content of \textit{Facebook} images posted by members of the 114th U.S. Congress (2015-2017) from 2012--2016. We use a convolutional neural network to automatically classify Republican or Democratic party affiliation solely from members' photographs. This model's predictions proxy political ideology along a left--right, liberal--conservative dimension, shown through comparisons of this measure with known Congress members' DW-NOMINATE ideology scores, the ``gold standard'' for political ideology estimation~\citep{poole2000congress}. 

% Would be good to be more specific here
In addition to this methodology, our measurement strategy allows us to shed light on liberal and conservative features of images. Content analysis of liberal and conservative images as predicted by our model reveals that clothing and depictions of the military are among the most predictive image elements distinguishing liberal from conservative images. 

Specifically, our paper aims to address three research questions. 
\begin{itemize}
  \item Can machine learning identify party affiliation and political ideology of politicians from their Facebook photographs?
  \item Can humans identify the party affiliation as well as a deep learning classifier? Do humans achieve a better accuracy or attend to the same visual features which the classifier utilizes? 
  \item  Which visual features are associated with liberals or conservatives? 
\end{itemize}

\section{Related Work}

Much of the prior work measuring political ideology has used voting data or text data extracted from political speeches.  Poole and Rosenthal pioneered a dimension reduction method known as the NOMINATE algorithm~\citep{poole2000congress}. The NOMINATE algorithm uses Members of Congress' roll call votes to score them along a left--right ideological continuum and is arguably one of the most important innovations in the political science literature of the 20th century. NOMINATE enables researchers to test and develop a broad swath of theories about political campaigns, polarization, parties and institutions~\citep{poole1984polarization,aldrich1995parties,krehbiel2010information,hix2002parliamentary}.  Ideological scores produced by the NOMINATE method through roll call data have become the ``gold standard'' against which  other means of measuring political ideology, including our own, are compared. 

Subsequent work measuring politicians' ideology shifted from voting data to text data. Earlier work in this area was mostly concerned with measuring ideology of different types of political actors using  political party platforms and politician's speeches~\citep{coffey2005measuring,sim2013measuring}. In more recent work, however, there are been greater interest in developing more general methods for measuring political ideology~\citep{grimmer2013text,iyyer2014political,sim2013measuring,slapin2008scaling}. Each new development provided a means of measuring political ideology for a wider variety of political actors and expanded the scope of information used to measure ideology, contributing to a richer picture of political ideology in its varying manifestations. 

For instance,~\citet{coffey2005measuring} estimates the ideological orientations of governors in the United States using ``state of the state'' speeches presented by governors on an annual basis. Moving beyond particular speeches to speeches in general,~\citet{sim2013measuring} develop a methodology for measuring ideology from any type of political speech. More recent work has developed an even more general means of measuring political ideology from texts, employing recursive neural networks and crowd-sourced data to estimate the political ideology of a single sentence~\citep{iyyer2014political}.  Ideology of Twitter users, politicians and not, is recoverable from who they follow \citep{Barbera2015}. 

Images, which provoke ideological dispositions even in non-political contexts~\citep{ahn2014nonpolitical},  have been largely ignored, mostly due to the difficulties of identifying and extracting their ideological content.  Discussions about the ideological content of images have been restricted primarily to the humanities~\citep{safran2001movie}. Recent advances in computer vision, however, allow one not only to make predictions about ideology from images, as we demonstrate below, but also to understand \textit{a priori} which \textit{aspects} of images reflect liberal and conservative ideological content, thereby providing a starting point for research exploring how political ideology is communicated through visual means \citep{joo2018image}.

\subsection{Motivation: Political Behaviors and Visual Media}
Images are a potent means of emotional persuasion which, with the advent of social media,  have become a routine form of political communication. While this development is sufficient motivation in itself for measuring the political ideology of images, there are more fundamental rationales, grounded in human cognition, for understanding how political ideology is communicated through images.

Kahneman (2016) identifies two primary neurological information processing systems: System 1 involves rapid, instinctual and emotional reactions to new information while System 2 involves slow, deliberative and logical reactions to new information. Psychological research exploring reactions to images, and visual stimuli more generally, consistently finds that images have a tendency to generate rapid, emotional reactions,  thereby suggesting that thoughts and opinions about the content of images are rooted in System 1 processes~\citep{turvey1973peripheral,pessoa2002attentional}.

In addition, recent studies adopt advanced computer vision methods in order to analyze large scale visual content data in political media, to overcome difficulty and cost in manual coding-based studies. 
\citet{joo2014visual} trains an automated visual classifier which can identify communicative intent of political images and assess emotional and professional portrayals of politicians, and \citet{huang2016inferring} uses deep learning to learn better visual features for the same task. 
%\citet{you2015multifaceted} also applies visual sentiment analysis to predict election outcomes using social media data.  
%\citet{Jo02018} shows how to use geolocated images to measure protest dynamics.
% WE CAN ADD THIS AFTER ACCEPTED

Political affiliation and ideology of politicians and supporters have been also studied by computational approaches. For example, \citet{joo2015automated} shows that Republican and Democratic politicians can be distinguished from their facial appearance by a hierarchical discriminative model. They annotated perceived personality and trait dimensions of politicians (\ie, trustworthy, attractive, competence, and so on) to train models to predict election outcomes and party affiliation of politicians. (See also~\citet{you2015multifaceted} for another computer vision based model for election prediction) Several methods have been proposed to automatically infer perceived (apparent) personality from visual cues ~\citep{ventura2017interpreting,junior2018first,escalera2018guest,escalante2018explaining}, and a recent study demonstrated that a similar automated method can reliably detect nonverbal behaviors of candidates during debates~\citep{joo2019ijoc}. \citet{wang2017polarized} also uses social media data to characterize conservative and liberal voters, using Twitter followers of Trump and Clinton in 2016 U.S. Presidential Election~\citep{wang2017gender}. 
Our paper differs from these works in that we focus on (1) \textit{\textbf{ideology}} rather than just binary party affiliations (\ie, ideology varies within the same party and is continuous.), and (2) systematically characterizing it on various dimensions beyond classification. 

These examples show that visuals are imperative to understand human behaviors, and computer vision based approaches can be applied to help decode various patterns of human social and political activities in order to study their impacts. Recently, scholars have adopted computer vision approaches and large scale visual data in social media for research projects in social science and media analysis~\citep{won2017protest,ha2018characterizing,zhang2018casm}. 
The main contribution of this paper is to demonstrate that computer vision methods can be used to explain and characterize political ideology from large scale social media data, scaling manual analysis currently conducted in the social sciences.   % Cites for this last sentence?

%joo2018social

%These works show the utility of computer vision as an efficient means for systematic measurement of visual communication. 

%If we conceive of images and text, then, as information delivery systems,  information delivered via textual means, which requires deliberation,  contrasts with similar information delivered via images in that the thoughts formed about textual information will be tend to be the product of more logical and deliberative, System 2 processes, while the thoughts formed about visual stimuli will tend to be the product of more emotionally based, System 1 processes. Thus, this study lays the groundwork for a deeper understanding of both the connection between political ideology and emotions and also how emotionally laden visual rhetoric is used by politicians to shape public opinion and voting behavior.

\section{Visual Framing of Political Ideology}

To understand how political ideology is projected through images, we must first understand how entities contained within images relate to political ideology. To accomplish this task, we attempt to understand the fundamental elements of political ideology and how basic features of images (objects, people, and events) reflect these elements.

\subsection{Fundamental elements of political ideology}
Ideologies  are sets of beliefs and values that allows individuals to simplify the complex political and policy world ~\citep{jost2009political,tetlock1983cognitive}. While some argue that political ideology is a multidimensional concept, the traditional definition of political ideology, which emerged during the Enlightenment in the 18th century, is that of a uni-dimensional, left--right continuum of ideological dispositions, later verified through a voluminous literature in social psychology and political science on the ideological orientations of mass publics~\citep{jost2009political}.

%While the conceptualization of political ideology is a hotly contested area of research within the social sciences, a relatively uncontroversial definition of ideology is presented by Erikson and Tedin (2003) as ``a set of beliefs about the proper order of society and how it can be achieved.''  

In the traditional conceptualization of left--right ideology of \citet{jost2009political},  ``conservative'' (right) and ``liberal'' (left) individuals are characterized as holding views which vary along two major dimensions: ``(1) advocating versus resisting social change'' and ``(2) rejecting versus accepting inequality.'' While those  on the right tend to be resistant to social change and accepting of inequality in its socioeconomic manifestations,  those on the left  tend to embrace social change and to reject inequality. These dimensions manifest as strong affinities toward nationalism,  capitalism, and status quo political and economic institutions~\citep{feldman2014understanding}. Similarly on the left, preferences favoring social change and a rejection of inequality have manifested as opposition to nationalism, sympathy for underprivileged groups in society and stronger preferences for redistributive policies. 

According to these definitions, right-leaning ideology should manifest in photos as objects or people, described in more detail below, that 1) suggest support for status quo political and economic institutions and; 2) suggest support for inequality. Conversely, left-leaning ideology should manifest in these image elements as  opposition to existing status quo political and economic institutions and opposition to inequality. Below we discuss each image element and provide some examples of how liberal and conservative ideological beliefs can be reflected withing each.

\subsection{Objects}
In childhood as early as 18 months old, humans recognize and use objects to convey symbolic meanings ~\citep{tomasello1999young}.
Indeed, the use of objects as signifiers of meaning in everyday life is so common that it generally goes unnoticed.
For example, flags are objects which represent nation--states, trees stand for environmentalism, hammers support the working class, and so on.
As we discuss above, political ideology can be reflected through objects which suggest support or opposition to  status quo political and economic institutions and to inequality.

\begin{figure}
\vspace{-15pt}
\centering
\includegraphics[width = .3\textwidth]{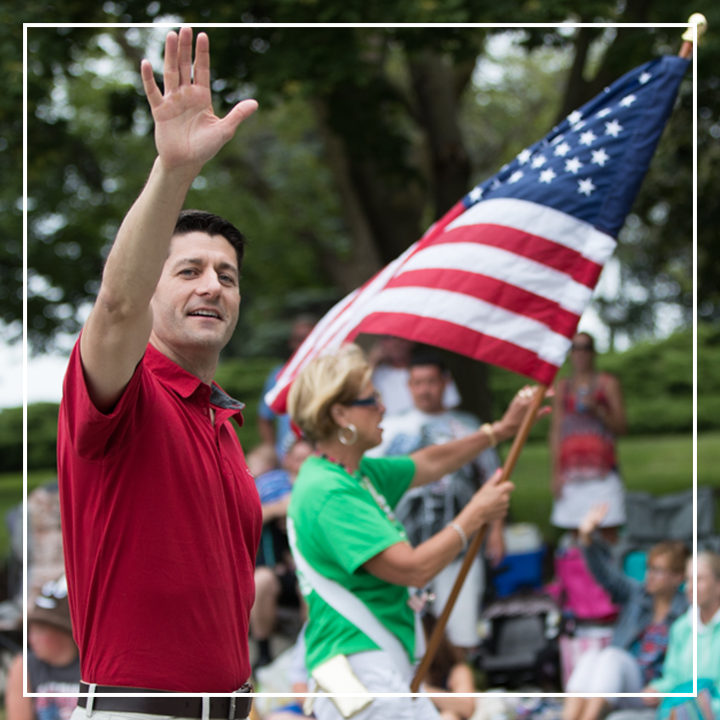}
\caption{Republican Speaker of the House of Representatives Paul Ryan addressing a crowd with an image of the American flag in the background. \textit{Source:} \url{https://www.facebook.com/speakerryan/} }
\label{fig:ob1}
\end{figure}

\begin{figure}
\vspace{-10pt}
\centering
\includegraphics[width = .3\textwidth]{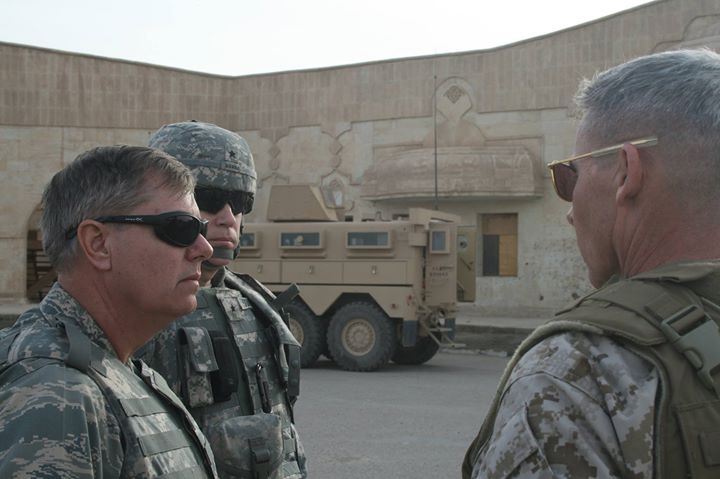}
\caption{Lindsey Graham, a  Republican Senator from South Carolina, meeting with members of the military \textit{Source:} \url{https://www.facebook.com/LindseyGrahamSC/}}
\label{fig:ob2}
\vspace{-10pt}
\end{figure}

Conservative ideology in the United States should be projected through objects that serve as symbols of nationalism, freedom, and capitalism while liberal ideology should be projected through objects that serve as symbols of inequality reduction. Thus we would expect that Republican members of Congress should have more photos that contain symbols of patriotism such as the American flag and symbols of military strength such as military equipment, as seen in Figures~\ref{fig:ob1} and~\ref{fig:ob2}.  % above from Republican members of Congress.  

\subsection{People}

\begin{figure}[!h]
	\centering
	\includegraphics[width = .3\textwidth]{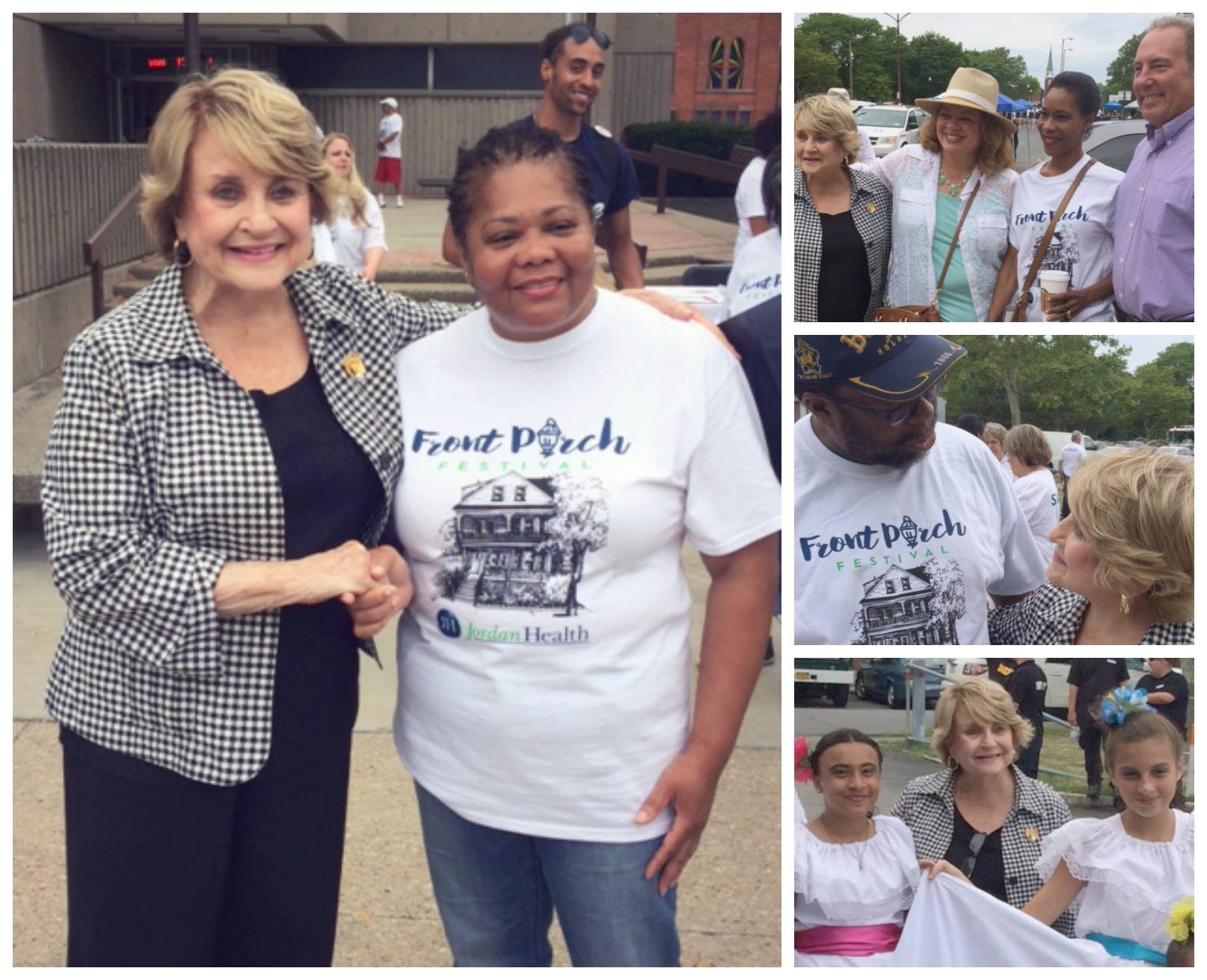}
	\caption{A photo collage posted on the Facebook profile of Represenative Louise Slaughter, a liberal Democrat (D-NY 25).  She is posing with African-American, Hispanic and white constituents at a festival in her district. \textit{Source:}\href{https://www.facebook.com/RepLouiseSlaughter/}{https://www.facebook.com/RepLouiseSlaughter/}.}
	\label{fig:p2}
\end{figure}

\begin{figure}[!h]
	\centering
	\includegraphics[width = .3\textwidth]{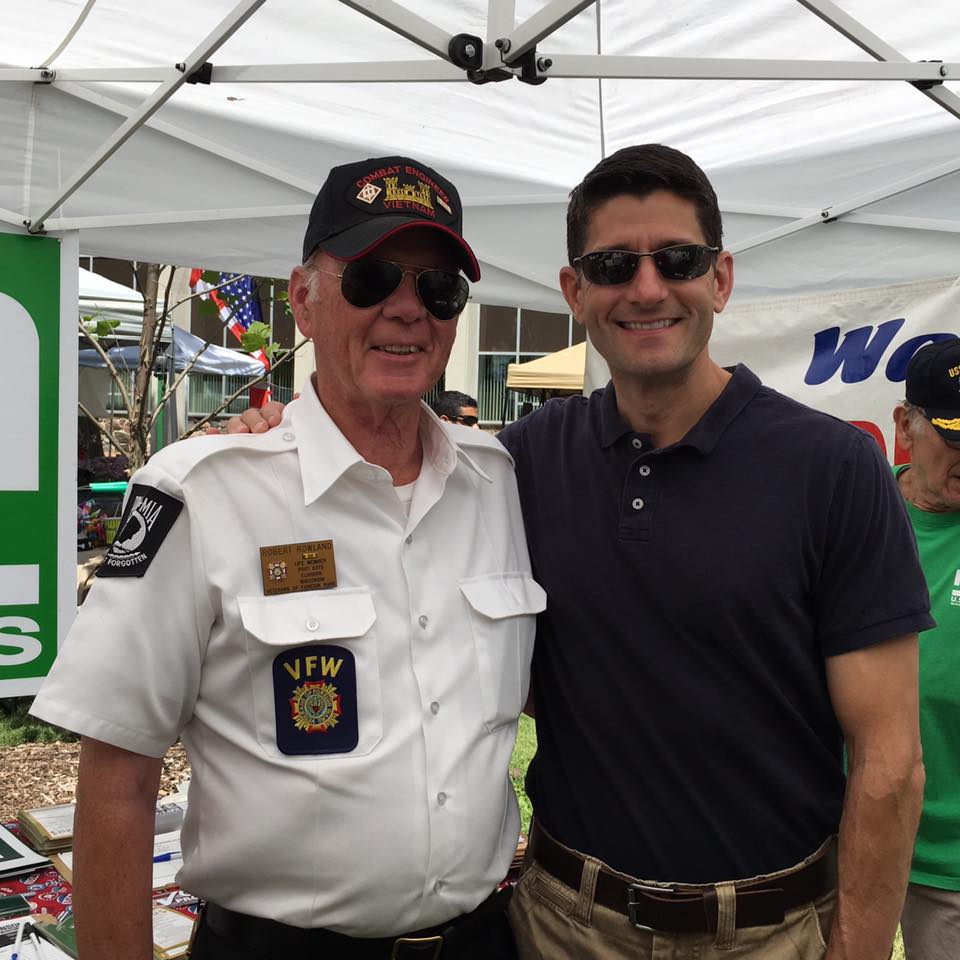}
	\caption{Republican House Speaker Paul Ryan Poses with a Vietnam Veteran. \textit{Source:} \href{https://www.facebook.com/paulryanwi/}{https://www.facebook.com/paulryanwi/}.}
	\label{fig:p1}
\end{figure}

People in images are relevant to political ideology to the extent that their inclusion in a photograph can reflect support or opposition to status quo political and economic institutions~\citep{Kreiss2019}. Thus, conservative politicians should be more likely to include individuals from dominant social groups such as white males, those who work in business, and members of the military, for example, while liberal politicians should be more likely to include members of under-represented minority groups, low wage workers, and protesters. Figure~\ref{fig:p1} presents an example of the former while Figure~\ref{fig:p2} presents an example of the latter.

\section{Data and Methods}
\subsection{Data: Facebook Photographs of Legislatures}
%\textbf{JASON: How did you collect the data?}
Facebook images of members of Congress were collected in March of 2016 over two weeks by a team of undergraduates and one of the authors.
%at the University of California, Berkeley. 
Using a YAML database of the social media accounts of current members of Congress at the time, a series of bots were written which identified the set of images from each member of Congress using their Facebook user id identified from the YAML file.\footnote{This database is constantly updated and maintained on GitHub by organizations such as GovTrack, ProPublica, MapLight and FiveThirtyEight and can be found here: \url{https://github.com/unitedstates/congress-legislators}}

These bots collected images along with associated image metadata. This process yielded  296,461 images for 319 Members of Congress. (As far as we are aware, this collection is the largest image database of Members of Congress.) Restrictions Facebook imposed during the collection process prevented us from collecting all 535 Members' photographs.

%\textbf{Below is how Marcus prepared subsets}
The initial dataset contains 319 politicians, 66 of them female and 55 non-white. As Democrats and Republicans have different gender and race ratios, using the entire dataset would lead to an unbalanced analysis, \eg, the classifier may be able to tell the party affiliation just based on the gender of politicians. To control for the effects of gender and race, we only use data from white male politicians in this paper. To make the analysis even more balanced, we chose 68 politicians from each party and randomly sampled 150 photographs from each politician after excluding accounts with fewer than 150 images. 

Our initial investigation of this dataset revealed that many images are not photographs but infographics, \eg, text or chart on clean background, and these types of visualizations are more frequently used by Democrats. While it is an interesting style of presentation, we exclude such infographics in our study by manually filtering them because they do not convey meaningful visual messages that we attempt to examine.

%In order to make the dataset more balanced
%Our image dataset was extracted from the Facebook profiles of 150 members of Congress, evenly balanced between Democrats and Republicans. From each profile, we randomly sampled 100 images and included them in our dataset. 
%Typical images include photos of the politician posing with veterans, children, and students, giving speeches, or attending outreach events and banquets (See Figures 3 and 4). However, many images did not include the politician; for example: sceneries, notices, documents, and graphics. For our purpose, graphics are posters or visual, usually with text, that advertise an event or cause. An example of a graphic can be seen in Figure X (INCLUDE IMAGE OF GRAPHIC). We found graphics to be particularly problematic in our analysis as our classifier relied too heavily on the difference between the usage of these graphics between political parties. Therefore, we manually removed them from our dataset using human annotators. 

%Additionally, we only included white male Congressmen in our dataset. This was to prevent our classifier from simply identifying women and people of color, of which significantly more politicians are liberal than conservative (CITE?). Our resultant dataset contained X (please fill) images: X politicians from each party and approximately X images per person. We randomly allocated our data into a 66-33 train-test split, with no politician appearing in both the training data and the testing data.

\subsection{Methods}
\subsubsection{Party Affiliation Classification}
In order to classify each image as belonging to a Democratic or Republican politician, we trained a convolutional neural network (CNN) with ResNet-34 architecture \citep{he2016deep} to take an image as input and generate a single output denoting the likelihood of being a Republican. We used a publicly available model provided by PyTorch pre-trained on ImageNet data and fine-tuned on our training set by Adam optimization \citep{kingma2014adam} with a learning rate of 0.0001.

\begin{figure*}[!h]
	\centering
	\includegraphics[width = .8\textwidth]{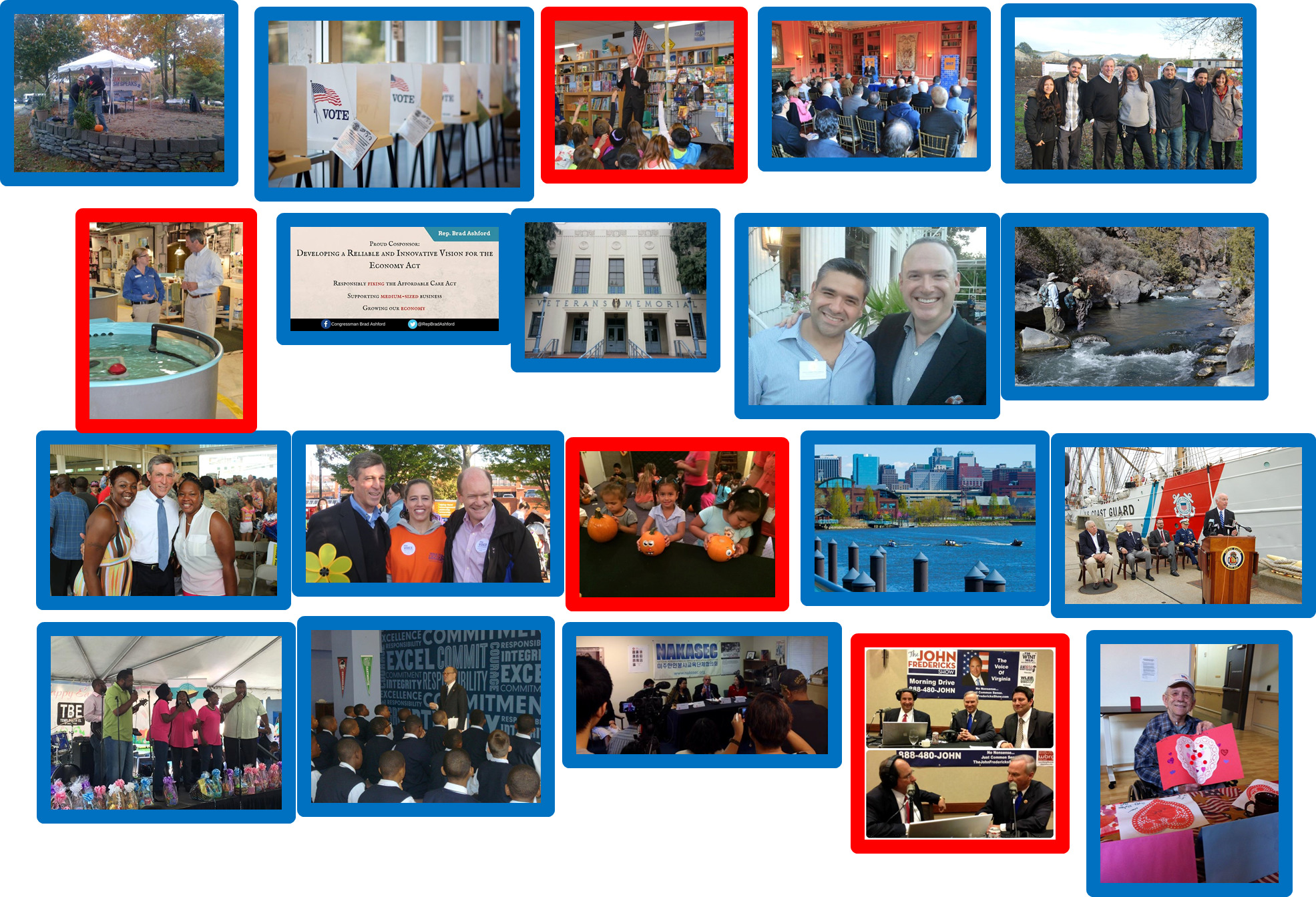}
	\caption{Top 20 predicted liberal images. The color of the bounding box shows the true party affiliation. Blue is Democrat and red is Republic.}
	\label{fig:top20}
\end{figure*}

\begin{figure*}[!h]
	\centering
	\includegraphics[width = .8\textwidth]{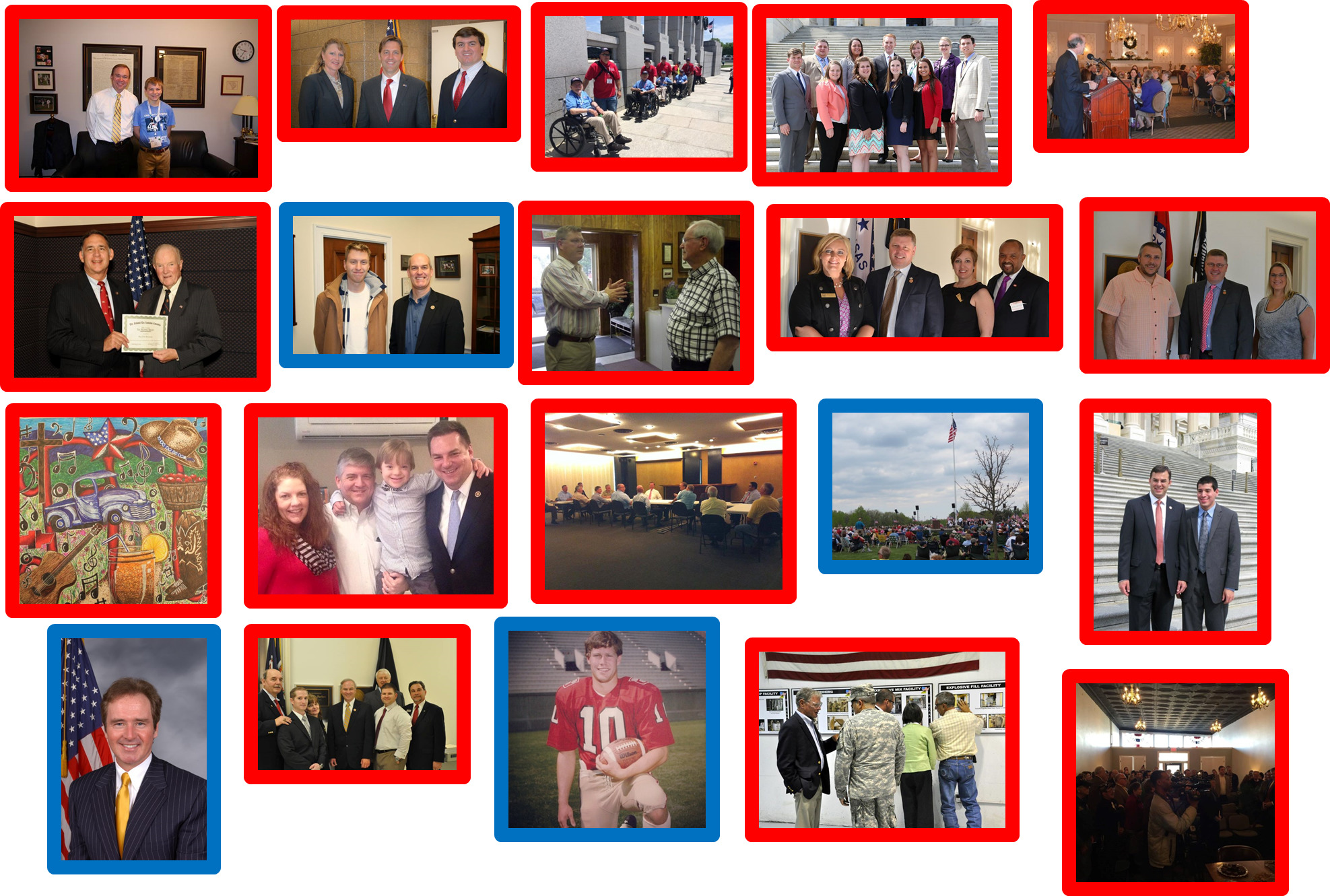}
	\caption{Top 20 predicted conservative images. The color of the bounding box shows the true party affiliation. Blue is Democrat and red is Republic.}
	\label{fig:bottom20}
\end{figure*}

% DO NOT FORGET THIS TABLE
We measured the classification accuracy through 10-fold cross validation.
%and report the accuracy in Table \ref{tab:pred}. 
In each run, we ensure each politician only appears either in the training set or the test set, but not both, by splitting the whole dataset by person. Splitting by person prevents the model from taking advantage of similarities of images taken in the same place or memorizing the facial appearance of specific individuals. The classifier correctly identifies the party affiliation of a politician with an average accuracy of 59.28\% (SD = 2.00). When we aggregate the classification scores from all of images for each politician and assign the average score to the person, a much higher accuracy, 82.35\%, is achieved. We show the most 20 liberal images and 20 most conservative images according to our classification scores in Figure~\ref{fig:top20} and Figure~\ref{fig:bottom20}.

%Moreover, we predicted the party affiliation of every politician by averaging prediction scores on his images (calculated on test set). Our model achieves 82.35\% prediction accuracy per politician. 

%It should note that we spitted the dataset into 10 folds by randomly assigning {\it politicians}, rather than images, to different folds. This guarantees that no images from same politicians exist in both training and test set. Therefore, the model is forced to learn the key features that can help identify the party affiliation, instead of memorizing the politicians. 

\subsubsection{Facial Expressions and Attributes}
While the above-mentioned classifier is trained on the whole image region to capture holistic visual features, we further analyze individual faces in photographs with a separate model. This is done by a two-step process. First, we distinguish the faces of the main politicians (\ie, the account owners) and the other people accompanying them in photographs (``associates''). Then, we apply CNNs to classify facial expressions, gender, and ethnicity. 

In the first step, we use an unsupervised learning approach to recognize the main politician in each image, because we do not have labeled facial photographs of politicians which can be used to train a face recognition model. Therefore, we first detect every face in each image and extract a 128-dimensional facial feature vector for each face using a pre-trained convolutional neural network provided by a computer vision package, dlib~\citep{king2009dlib}. Then we cluster these feature vectors for all faces detected from all images posted by each politician. We use Chinese-Whispers clustering \citep{biemann2006chinese}, an agglomerative clustering method that does not require a predefined number of clusters, on the 128-dimension representations, treating the faces in the largest cluster as the politician's faces. We have manually validated that this method is very accurate to separate the main politicians' faces. 

%Since the face images are mixed with politicians and non-politicians, we first apply the clustering method to separate the two groups. We extract all faces from each politician's Facebook profile and take them as the input of a  pre-trained deep convolutional neural network \citep{king2009dlib}. The output of the neural network is a 128-dimensional vector which we use as the representation of the input face. Then we conduct Chinese-Whispers clustering \citep{biemann2006chinese}, an agglomerative clustering method that does not require a predefined number of clusters, on the 128-dimension representations, treating the faces in the largest cluster as the politician's faces. 

%In order to detect if a politician's face is clustered into more than one cluster due to the high variance of that person's face images, we train a logistic regression classifier to find the false positive and false negatives in our separation result. To train the logistic regression, we treat all faces from the largest cluster as positive samples and randomly select a large amount of faces from other clusters as negative samples (10 times as positive samples). The binary classifier detects a very small number of false positive and false negative faces. We manually check those faces and find most of them are difficult to tell whether they are politicians or non-politicians. %This illustrates that our clustering methods can correctly separate politicians and non-politicians from their face images.   

\begin{figure}[!h]
	\centering
	\includegraphics[width = .4\textwidth]{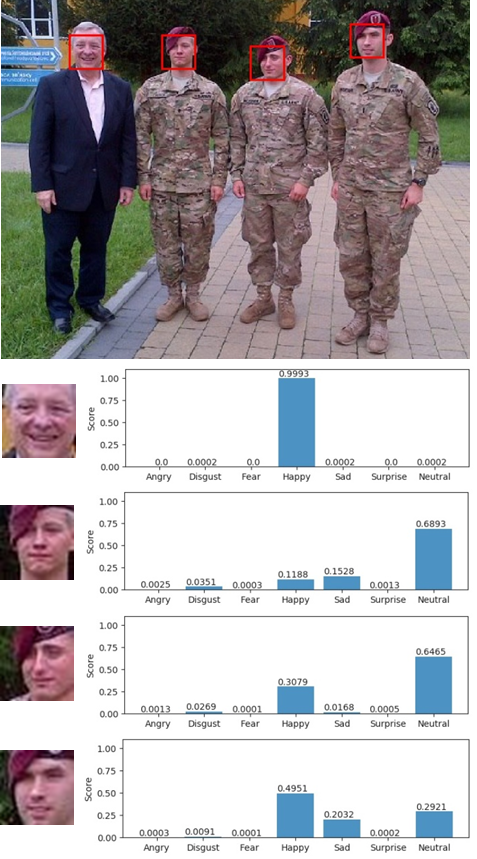}
	\caption{Predicting Facial Expression}
	\label{fig:expression}
\end{figure}

In the second step, we identify the face expression, gender, and ethnicity of politicians and associates from their images by three separate CNNs.  These CNNs are trained on three public face image datasets respectively, i.e., Facial Expression in the Wild (ExpW) \citep{SOCIALRELATION_2017}, CelebFaces Attributes (CelebA) \citep{liu2015faceattributes}, and UTKFace~\citep{zhifei2017cvpr}. ExpW contains 106,962 face images, with each labeled by one of seven expression categories: angry, disgust, fear, happy, sad , surprise, and neutral. CelebA contains 202,599 faces images and 40 binary attributes annotations per image. UTKFace contains 24,108 faces with ethnicity annotations, including White, Black, Asian, Indian, and others.\footnote{The `others' category combines multiple races such as Hispanic and Middle Eastern Asian but does not distinguish the subcategories. } We choose ResNet-34~\citep{he2016deep} as the network architecture, and the classification accuracy on the three datasets is 73.19\%,\footnote{This is a multiclass classification and the chance accuracy is 14\%.} 91.47\%, and 86.19\%, respectively. 
%We apply the trained models on every Facebook image to extract the face expression of politicians and non-politicians, and the face expression, ethnicity and ethnicity of non-politicians. 
Figure~\ref{fig:expression} shows one example of the prediction on face expression.

We calculated the summary statistics of predicted face expressions on politicians and non-politicians in Table~\ref{tab:expression_p} and Table~\ref{tab:expression_np}. We also conduct two-proportion Z test on all expression categories, which shows that for politicians, all expression categories (except for 'Fear') are significantly different in Democrats and Republicans. Meanwhile, for non-politicians in politicians' images, `Happy', `Surprise', and `Neutral' are significantly different between parties. Following the same idea, we calculate the summary statistics of predicted race and gender of non-politicians in politicians' images. The result is summarized in Table~\ref{tab:race_np}. Democrats tend to have more minorities and fewer man in their images, and the difference is significant.

\subsubsection{Objects and Events}
Our holistic classifier is trained on the whole image region and thus able to leverage object and scene features. However, since we did not train the model with object or event labels, it is difficult to understand which objects or events are associated with the party affiliation and thus play an important role in classification.  To better identify important individual semantic features, we use the Google Vision API, as it provides meaningful labels for the images in our dataset.  

In the computer vision literature, there exist a number of dataset choices that can be used for object detection or classification, such as PASCAL-VOC~\citep{everingham2010pascal}, MS-COCO~\citep{lin2014microsoft}, or Imagenet~\citep{russakovsky2015imagenet}. We did not use these existing datasets and trained our own models because the categories defined in these datasets are limited and not strongly related to political or social dimensions of human activities. 

%That API returns a machine-generated identifier (MID) corresponding to the entity's Google Knowledge Graph entry, a description attribute which describes the detected label, and a [0-1] confidence score. Google provided 2,285 labels, 1,874 for the Democrats and 1,758 for Republicans, for our images.

%which hides the complexity machine learning models as well as image processing algorithms and can quickly execute Image Content Analysis on the entire image dataset and summarize results into relevant labels (keywords, objects and categories). Labels are selected among thousands of object categories and mapped to the official Google Knowledge Graph. By passing the request with image content and features type, Google Vision API will perform the image content analysis on the cloud and return a JSON like structure reply including the label annotations. Each annotation will contain a mid attribute, which is the machine-generated identifier (MID) corresponding to the entity's Google Knowledge Graph entry, a description attribute which describes the detected label shortly and a score that shows the confidence which ranges from 0 (no confidence) to 1 (very high confidence). We have detected 2285 different labels from all the images among which Dem owns 1874 and Rep owns 1758. 

%\begin{figure}[!h]
%	\centering
%	\includegraphics[width = .4\textwidth]{./figs/hist1.png}
%	\caption{Top 50 most frequent labels distribution on Dem and Rep \footnote{\textit{Source:} 
%	\href{}.}}
%	\label{fig:p1}
%\end{figure}

\begin{figure}
    \centering
    \includegraphics[width=0.5\textwidth]{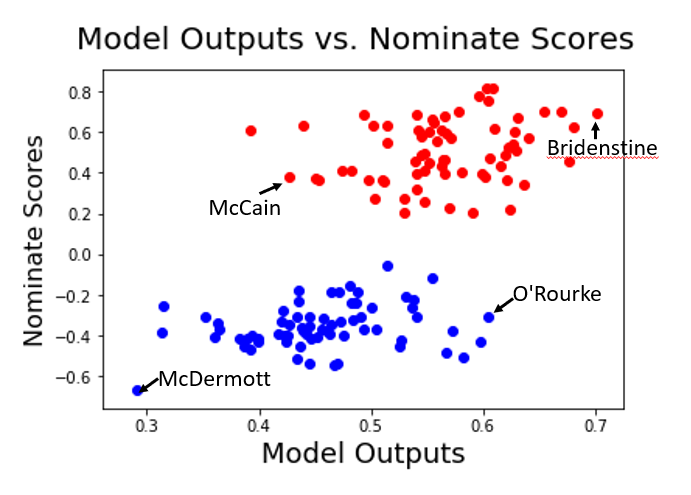}
    \caption{Correlation between gold standard NOMINATE scores and predicted political ideology of the image (r = 0.6736 and p-value = 7.794e-25).}
    \label{fig:nominate}
\end{figure}

\section{Results}
%As we mention above, the ultimate goal of this paper is to measure political ideology within images using image data from members of Congress. After training the classifier to identify images which originated from Democratic or Republican members of Congress, we sought to determine whether 

%The predicted probabilities generated from our model can be utilized to measure the ideological content of images.  Figure \ref{fig:nominate} shows this main result.

%when compared to the ``gold standard'' DW-NOMINATE ideological scores mentioned above. 

%ZST: I got rid of "DW-" because we say "NOMINATE" everywhere else.  I know they are slightly different, but no point confusing the reader.

Figure~\ref{fig:nominate} shows the most recent NOMINATE scores of Members of Congress with the average classification score obtained all images posted by each person.  The correlation between NOMINATE scores and the average classification score generated by the model is very high ($r=0.6736$). For inner-party correlation between NOMINATE scores and model output, we have ($r=0.2595$) for the Democratic and ($r=0.2314$) for the Republican. %This result is particularly striking because NOMINATE scores are estimated using the voting behavior of Members of Congress while our measure is generated entirely from images that they post on Facebook.
%Note that we only use the NOMINATE scores for a validation purpose and do not use them in training. 
This result confirms that the political ideology solely estimated from images is indeed correlated with the known ideology inferred from actual voting data. 

%These results provide strong evidence that our method allows us measure the ideology of members of Congress from their images. In the subsequent analyses below, we use the measure of image ideology that we developed to understand the features of images that distinguish ``liberal'' and ``conservative'' images. 

\subsection{Elements of Image Ideology}
\begin{figure}
    \centering
    \includegraphics[width = .47\textwidth]{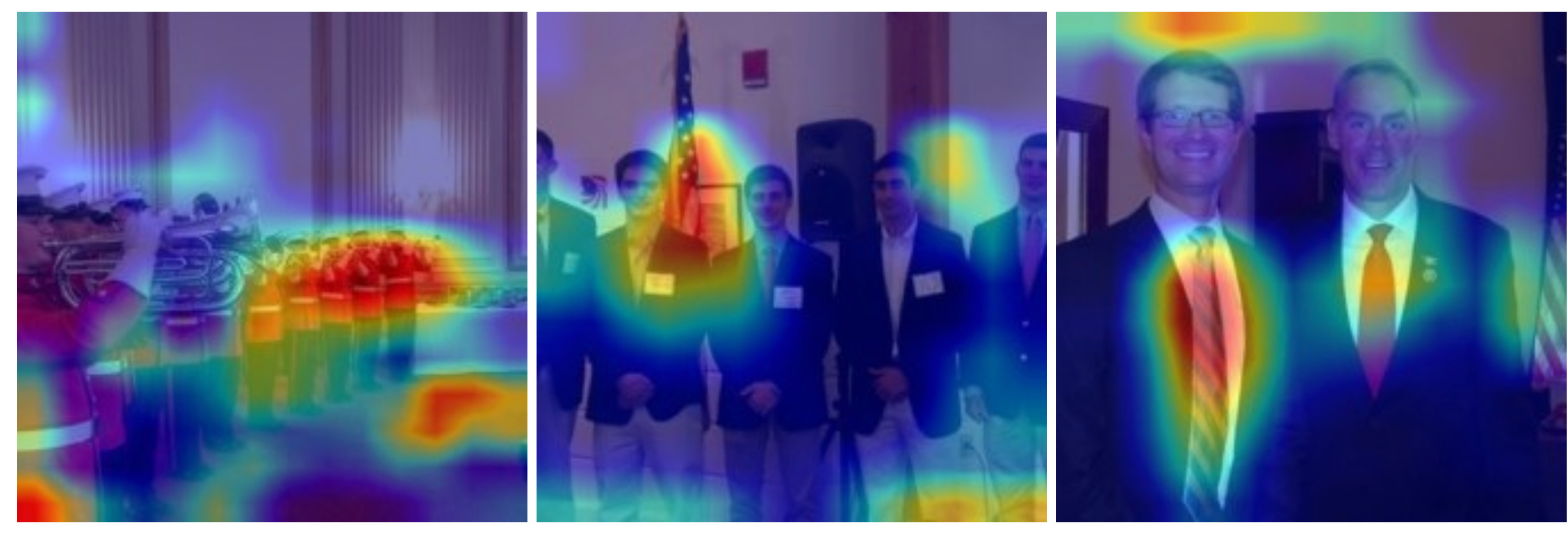}
    \caption{Salient features within images for predicting conservative images using gradient based localization (Grad-CAM)}
    \label{fig:gcon}
\end{figure}

\begin{figure}
    \centering
    \includegraphics[width = .47\textwidth]{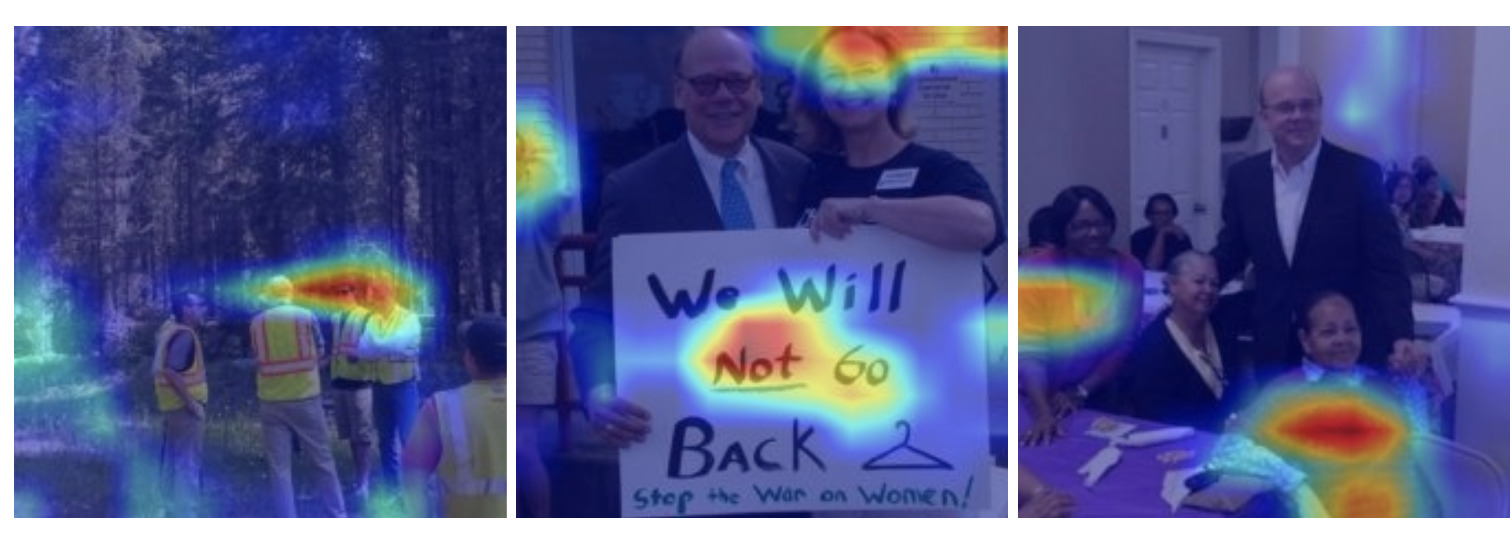}
    \caption{Salient features within images for predicting liberal images using gradient based localization (Grad-CAM)}
    \label{fig:glib}
\end{figure}

Using the scores generated by our model, we further explore image features that distinguish liberal from conservative images. Initial analyses using gradient based localization (Grad-CAM)~\citep{selvaraju2017grad} produce interesting results, as seen in Figures~\ref{fig:gcon} and~\ref{fig:glib}. For instance, when predicting conservative images, salient image features are those which correspond to maintenance of status quo economic and political institutions such as the military (military band, Figure~\ref{fig:gcon}, left), patriotism (American flag, Figure~\ref{fig:gcon}, center)  and business (ties, Figure~\ref{fig:gcon}, right).

Salient image features predicting liberal images, on the other hand, appear to be related to concerns about economic inequality and members of minority groups. In Figure~\ref{fig:glib}, for instance, the most salient aspects of liberal images detected are the hard hats of construction workers (left), a female and a protest sign (center), and two individuals who are members of minority groups (right).

% \begin{figure}[ht!]
%     \centering
%     \includegraphics[width = .5\textwidth]{./figs/conimgelements}
%     \caption{Top 20 conservative/Republican image elements identified by the Google API.}
%     \label{fig:conelts}
% \end{figure}

% \begin{figure}[ht!]
%     \centering
%     \includegraphics[width = .5\textwidth]{./figs/liberalimgelment}
%     \caption{Top 20 liberal/Democratic image elements identified by the Google API.}
%     \label{fig:libelts}
% \end{figure}

% \begin{figure}[ht!]
%     \centering
%     \includegraphics[width = .5\textwidth]{./figs/positive_corr}
%     \caption{Top 30 image elements identified by the Google API sorted by P-Value with positive correlation.}
%     \label{fig:corrPos}
% \end{figure}

\begin{table*}[!htb]
    \caption{ Image labels identified by Google Vision API. \% (D) and \% (R) represent the proportion of images containing each label in Democratic and Republican parties.}
    \begin{subtable}{.5\linewidth}
      \centering
        \caption{Top 30 Democrat-associated labels}
\begin{tabular}{ccccc}
\hline
\textbf{Label} & \textbf{$\chi^2$} & \textbf{p-value}& \textbf{\% (D)} & \textbf{\% (R)}\\ \hline
sea               &  58.7 &  1.85E-14 &          .013 &   .0029 \\
shore             &  45.7 &  1.35E-11 &      .009 &   .0015 \\
coast             &  44.2 &  3.01E-11 &           .008 &   .0011 \\
ocean             &  41.7 &  1.09E-10 &           .007 &   .0009 \\
line              &  36.6 &  1.44E-09 &           .007 &   .0012 \\
nature reserve    &  35.8 &  2.20E-09 &          .010 &   .0025 \\
tree              &  34.1 &  5.14E-09 &           .113 &   .0865 \\
oceanic view &  33.5 &  7.29E-09 &          .006 &   .0007 \\
promontory        &  30.3 &  3.65E-08 &          .004 &   .0000 \\
nature            &  30.3 &  3.79E-08 &          .009 &   .0026 \\
horizon           &  29.9 &  4.50E-08 &          .009 &   .0028 \\
leaf              &  29.7 &  5.13E-08 &           .007 &   .0017 \\
sky               &  29.5 &  5.45E-08 &           .049 &   .0331 \\
reflection        &  26.6 &  2.56E-07 &          .009 &   .0027 \\
headland          &  26.3 &  2.94E-07 &         .003 &   .0000 \\
wilderness        &  25.2 &  5.28E-07 &         .007 &   .0020 \\
autumn            &  22.1 &  2.60E-06 &       .005 &   .0008 \\
sunset            &  21.3 &  4.01E-06 &         .005 &   .0010 \\
area              &  20.5 &  5.97E-06 &         .006 &   .0017 \\
water             &  19.9 &  8.26E-06 &         .026 &   .0159 \\
ecosystem         &  18.2 &  1.97E-05 &         .009 &   .0036 \\
beach             &  17.6 &  2.74E-05 &        .004 &   .0006 \\
metropolitan  &  16.4 &  5.01E-05 &         .005 &   .0012 \\
home              &  16.2 &  5.60E-05 &       .008 &   .0032 \\
vegetation        &  16.1 &  6.14E-05 &         .006 &   .0018 \\
calm              &  14.9 &  1.14E-04 &        .003 &   .0005 \\
sunrise           &  14.8 &  1.18E-04 &          .004 &   .0012 \\
mountain          &  14.3 &  1.53E-04 &      .006 &   .0023 \\
education         &  13.8 &  2.01E-04 &         .015 &   .0089 \\ 
body of water     &  13.5 &  2.39E-04 &       .004 &   .0014 \\ \hline
\end{tabular}
\label{tab:corr_pos}
    \end{subtable}%
    \begin{subtable}{.5\linewidth}
      \centering
        \caption{Top 30 Republican-associated labels}
\begin{tabular}{cccccc}
\hline
\textbf{Label} & \textbf{$\chi^2$} & \textbf{p-value} &  \textbf{\% (D)} & \textbf{\% (R)}\\ \hline
military officer &  38.4 &  5.83E-10 &       .0382 &   .0583 \\
military person  &  30.5 &  3.40E-08 &       .0340 &   .0509 \\
award            &  26.1 &  3.16E-07 &      .0418 &   .0587 \\
official         &  25.9 &  3.67E-07 &       .2293 &   .2623 \\
uniform          &  25.5 &  4.43E-07 &      .0134 &   .0238 \\
troop            &  22.1 &  2.58E-06 &      .0162 &   .0266 \\
standing         &  19.2 &  1.15E-05 &      .0095 &   .0172 \\
wheelchair       &  18.1 &  2.14E-05 &     .0007 &   .0038 \\
senior citizen   &  16.4 &  5.05E-05 &     .1014 &   .1207 \\
court            &  14.1 &  1.78E-04 &      .0033 &   .0075 \\
banquet          &  13.1 &  2.92E-04 &        .0048 &   .0095 \\
military uniform &  12.4 &  4.41E-04 &     .0076 &   .0131 \\
soldier          &  12.2 &  4.89E-04 &       .0120 &   .0186 \\
military rank    &  11.8 &  5.98E-04 &      .0062 &   .0111 \\
function hall    &  11.7 &  6.31E-04 &      .0070 &   .0122 \\
conference hall  &  11.4 &  7.22E-04 &      .0125 &   .0189 \\
picture frame    &  11.2 &  8.22E-04 &     .0025 &   .0059 \\
artwork          &  10.5 &  1.18E-03 &      .0061 &   .0107 \\
outerwear        &  10.3 &  1.35E-03 &       .0171 &   .0241 \\
community        &  10.3 &  1.35E-03 &      .1365 &   .1536 \\
army             &  10.1 &  1.47E-03 &      .0162 &   .0230 \\
formal wear      &   9.4 &  2.17E-03 &    .0446 &   .0547 \\
visual arts      &   9.3 &  2.32E-03 &        .0031 &   .0063 \\
team             &   8.7 &  3.15E-03 &       .0542 &   .0649 \\
military         &   8.7 &  3.24E-03 &    .0180 &   .0245 \\
livestock        &   8.6 &  3.43E-03 &       .0001 &   .0015 \\
shoulder         &   8.4 &  3.77E-03 &       .0042 &   .0077 \\
windshield       &   8.0 &  4.59E-03 &      .0000 &   .0011 \\
grandparent      &   7.9 &  4.85E-03 &      .0035 &   .0066 \\
tours            &   7.9 &  4.91E-03 &       .0039 &   .0071 \\ \hline
\end{tabular}
\label{tab:corr_neg}
    \end{subtable} 
\end{table*}

To further explore differences between conservative and liberal images, we use the visual labels detected by the Google Vision API and report two results. First, Table~\ref{tab:corr_pos} and~\ref{tab:corr_neg} show the detected labels sorted by their relevance to the party affiliation using $\chi^2$ statistics. These statistics are computed for each label independently. 
%In addition, we also use predicted probabilities of the images generated by our classifier to further classify a group of the test images into ``conservative'' ($\mathbb{P}[Republican] \geq 0.90$) and ``liberal'' ($\mathbb{P}[Republican] \leq 0.10$) images.  Logistic regression then identifies image labels that best predict these respective groups. Figures~\ref{fig:conelts} and~\ref{fig:libelts} show the 20 most predictive image labels, arranged by p-value. 

%In addition to this we also explore image elements that are correlated with liberal/Democratic and conservative/Republican images more generally (Figure~\ref{fig:corrPos} and~\ref{fig:corrNeg}). 

Our findings from these analyses reveal a great deal about how  political ideology is reflected through images. Conservative members of Congress tend to use image objects and people to project support for status quo political institutions and hierarchy maintenance such as the military (``military officer'', ``military person'', ``uniform'', ``troop'') and objects which signify state and economic power (``court'', ``formal wear'') while liberal members of Congress seek to project an image of care for the global community (``metropolitan area'', ``education'') and environmental protection (``ocean'',``nature reserve'', ``wilderness'').

Differences between image elements in each of these photos also correspond to work on the ``Big Five'' personality differences discovered between liberals and conservatives~\citep{carney2008secret}. Liberals were found to be high in ``Openness'' and low in ``Conscientiousness'' while conservatives were found to possess the opposite character traits, being low in ``Openness'' but high in ``Conscientiousness.'' According to \citet{carney2008secret}, traits of people high in Openness include ``life‐loving, free, unpredictable'' and ``open to experience'' while traits of people high in Conscientiousness include ``definite, persistent, tenacious'', ``tough, masculine, firm'' and ``reliable, trustworthy, faithful, loyal.'' These high Conscientiousness traits clearly correspond strongly to values such as emphasized in the military which are present in conservative images while high Openness traits can arguably be symbolized by expansive natural scenes such as oceans, the sky and so on. 

%Figure~\ref{fig:conid} and ~\ref{fig:libid} show how the top 20 image elements for liberal and conservative images relate to support or opposition to the status quo and inequality. Many of the predictive image elements for each group are neutral along these dimensions.  In conservative images in particular, however, objects and people which signify support for the status quo and inequality are prominent. 

%\begin{figure}[h!]
%    \centering
 %   \includegraphics[width = .5\textwidth]{./figs/repimgdims}
%    \caption{Image elements from conservative along status quo and inequality dimensions of political ideology.}
%    \label{fig:conid}
%\end{figure}

%\begin{figure}[h!]
 %   \centering
%    \includegraphics[width = .5\textwidth]{./figs/demimgdims}
 %   \caption{Image elements from liberal images along status quo and inequality dimensions of political ideology.}
 %   \label{fig:libid}
%\end{figure}

%Specifically, we can see from Figure~\ref{fig:repimgdims} that the image elements aligned most distinctly with conservative ideology are those that can arguably signify support for the status quo and support for inequality. For instance, military rank, competition and formal wear arguably suggest support for the status quo and support for inequality.

%One more general conclusion that might be drawn from this finding is that, since Republican members of Congress attempt to appeal to a broad audience on Facebook, they tend to use more subtle means of conveying political ideology through their attire rather than more blatant symbols such as flags, weapons and so on. 

\subsection{Emotions and Diversity}
We now turn to emotional expressions of politicians in Table~\ref{tab:expression_p} and associates in Table~\ref{tab:expression_np}. We find that in both cases Republicans tend to express happiness (smile) more in their photos than Democrats. Our finding is consistent between politicians and their associates, who are more likely to share the same ideological stances as the politicians they accompany. Indeed, recent surveys conducted in the U.S. suggest that conservatives are happier and more satisfied with their lives than liberals based on self-reported data~\citep{taylor2006we, carroll2007most}. %Our finding is consistent with these survey data while we measure ``displayed'' happiness from Facebook photographs. 

In contrast, our finding directly contradicts a recent study published in \textit{Science}~\citep{wojcik2015conservatives} in which the authors argue that liberals display greater happiness based on analysis closely related to ours. In their study, the authors analyzed smiling behavior of congressmen from the Congressional Pictorial Directory of the 113th U.S. Congress and reported that smiling is negatively correlated with political conservatism. 
One possible reason for this discrepancy is data. The official profile photographs of congressmen most likely exhibit ``standard'' smiling expression with very few exceptions.  Our study uses Facebook photographs that depict more types of activities in much larger quantity than than Congressional Pictorial Directory. 

Finally, we find that Democratic politicians post photographs with more people from non-white groups and more females, as shown in Table~\ref{tab:race_np}. This is an expected result as discussed earlier.\footnote{We found the `Indian' category was sometimes triggered by Hispanic faces, which may explain high proportions of Indian in both parties.}

%These findings are interesting because they are supported by research which finds that conservatives tend to report being happier more generally, but differ from results reported in \textit{Science}~\citep{wojcik2015conservatives} based on analyses of a smaller set of photographs which suggest that liberals tend to smile more. 

% expression of politician table
\begin{table}[]
\small
\caption{Face expressions of main politicians}
\begin{tabular}{ccccc}
\hline
\textbf{Expression} & \textbf{Dem (\%)} & \textbf{Rep (\%)} & \textbf{Dem-Rep (\%)} & \textbf{p-value} \\ \hline
Angry      & 0.305    & 0.029    & 0.276        & \textbf{*0.0052} \\ %\hline
Disgust    & 0.244    & 0        & 0.244        & \textbf{*0.0038} \\ %\hline
Fear       & 0        & 0        & 0            & NaN              \\ %\hline
Happy      & 65.365   & 71.412   & -6.047       & \textbf{*0.0000} \\ %\hline
Sad        & 1.404    & 0.757    & 0.647        & \textbf{*0.0010} \\ %\hline
Surprise   & 1.861    & 0.873    & 0.988        & \textbf{*0.0005} \\ %\hline
Neutral    & 30.821   & 26.929   & 3.892        & \textbf{*0.0004} \\ \hline
\end{tabular}
\label{tab:expression_p}
\end{table}

% expression of non-politician table
\begin{table}[]
\small
\caption{Face expressions of associates}
\begin{tabular}{ccccc}
\hline
\textbf{Expression} & \textbf{Dem (\%)} & \textbf{Rep (\%)} & \textbf{Dem-Rep (\%)} & \textbf{p-value} \\ \hline
Angry               & 0.079             & 0.109             & -0.030                & 0.2735           \\ %\hline
Disgust             & 0                 & 0                 & 0.244                 & 0.4037           \\ %\hline
Fear                & 0.009             & 0             & 0.009                     & 0.3169                \\ %\hline
Happy               & 62.688            & 66.502            & -3.814                & \textbf{*0.0000} \\ %\hline
Sad                 & 1.118             & 1.138             & -0.020                & 0.0683           \\ %\hline
Surprise            & 0.554             & 0.436             & 0.118                 & \textbf{*0.0237} \\ %\hline
Neutral             & 35.550            & 31.806            & 3.744                 & \textbf{*0.0000} \\ \hline
\end{tabular}
\label{tab:expression_np}
\end{table}

% race of non-politician table
\begin{table}[]
\small
\caption{Race and gender distribution of associates.}
\begin{tabular}{ccccc}
\hline
\textbf{Race/Gender} & \textbf{Dem (\%)} & \textbf{Rep (\%)} & \textbf{Dem-Rep (\%)} & \textbf{p-value} \\ \hline
White         & 71.219            & 82.876            & -11.657               & \textbf{*0.0000} \\ %\hline
Black         & 12.818            & 6.419             & 6.399                 & \textbf{*0.0000} \\ %\hline
Asian         & 4.658             & 3.190             & 1.468                 & \textbf{*0.0292} \\ %\hline
Indian        & 8.723             & 5.876             & 2.847                 & \textbf{*0.0000} \\ %\hline
Others        & 2.581             & 1.639             & 0.942                 & \textbf{*0.0055} \\ \hline
Male        & 63.040             & 64.487             & -1.447                 & \textbf{*0.0322} \\ \hline
\end{tabular}
\label{tab:race_np}
\end{table}

\subsection{Comparison Against Human Perception}
How do humans conceptualize ideology? Will it be different from what our classifier has learned from the data? To answer these question, we performed asked participants to guess the party affiliation of politicians given their Facebook photographs and compared their accuracy with our model classification. We first randomly sampled 2 images for each of 136 male politicians in out dataset, leading to 272 images in total. Participants were recruited from Amazon Mechanical Turk with the location limited to the U.S., and each annotator was given only one image at a time. We asked them if they recognized the target politician, and if so, discarded their responses. Each of the 272 images was rated by 50 annotators, and the party for each image was determined by the majority voting. 3 images received the same number of votes for both parties, so an additional annotator was added to break the ties. 

The classification accuracy of human judgment was 61.0\%, and our model yielded 63.6\% on the same set.  This difference is not statistically significant based on McNemar's test. The annotators tended to choose Republican (58.1\%) more than Democrat (41.9\%) in their responses (Table~\ref{tab:human-comp} and Table~\ref{tab:model-comp}), and therefore their responses were more accurate in the true-Republican subset. This imbalance may be related to the fact that our dataset only contains male politicians. 

\begin{figure*}
    \centering
    \includegraphics[width = .6\textwidth]{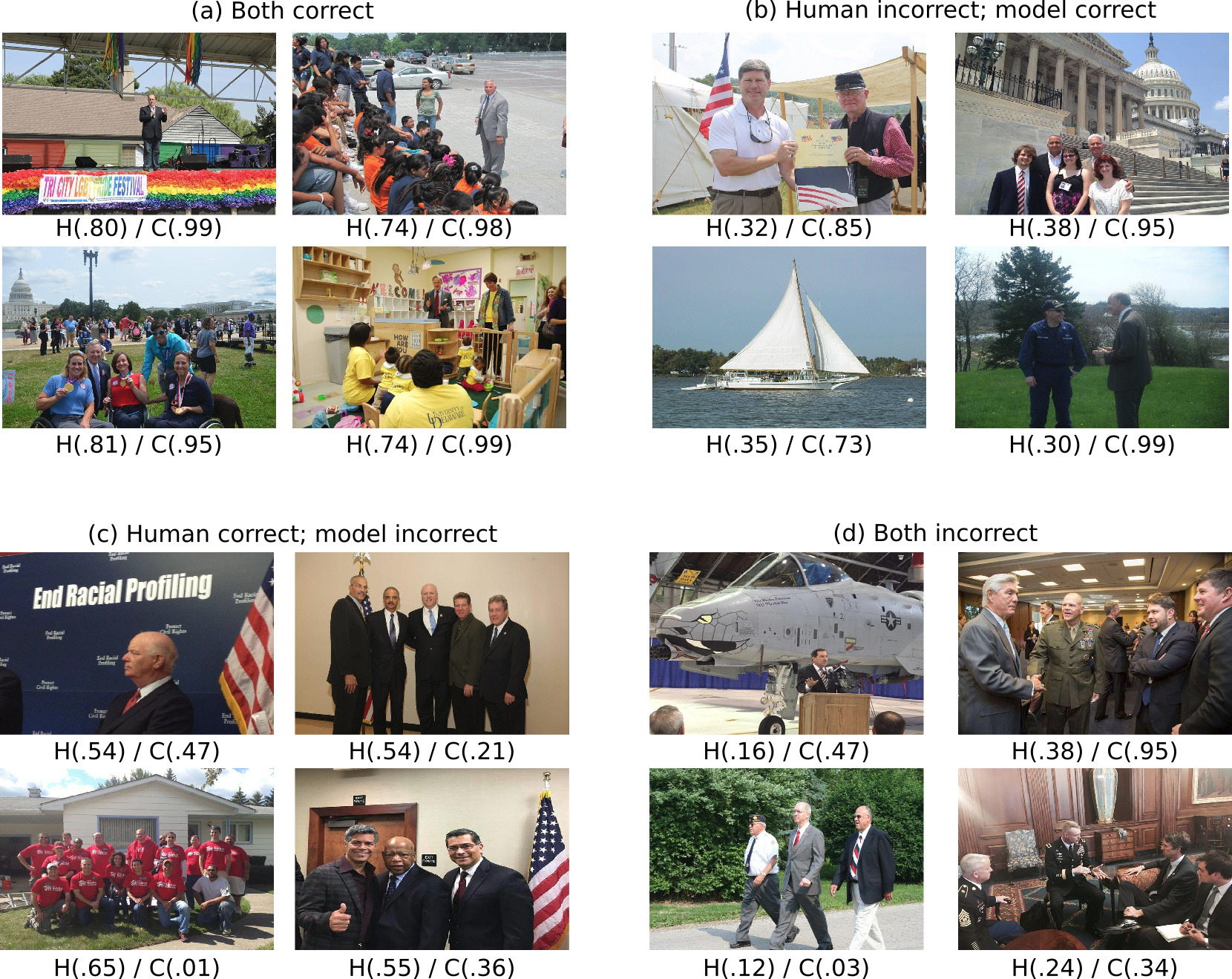}
    \caption{Comparison of party classifications by human judges and our model for \textbf{Democrats}. A higher score indicates Democrat (H: Human, C: Classifier). (a) Both identified key features correctly (\eg, children, LGBT). (b) Humans made wrong associations (\eg, yacht = rich). (c) The model could not infer true meanings of symbols (\eg red t-shirts) or recognize text (\eg, ``End Racial Profilling'') or people. (d) Some Democrats exhibit features and activities similar to Republicans (\eg military officers). }
    \label{fig:dem-comp}
\end{figure*}

\begin{figure*}
    \centering
    \includegraphics[width = .6\textwidth]{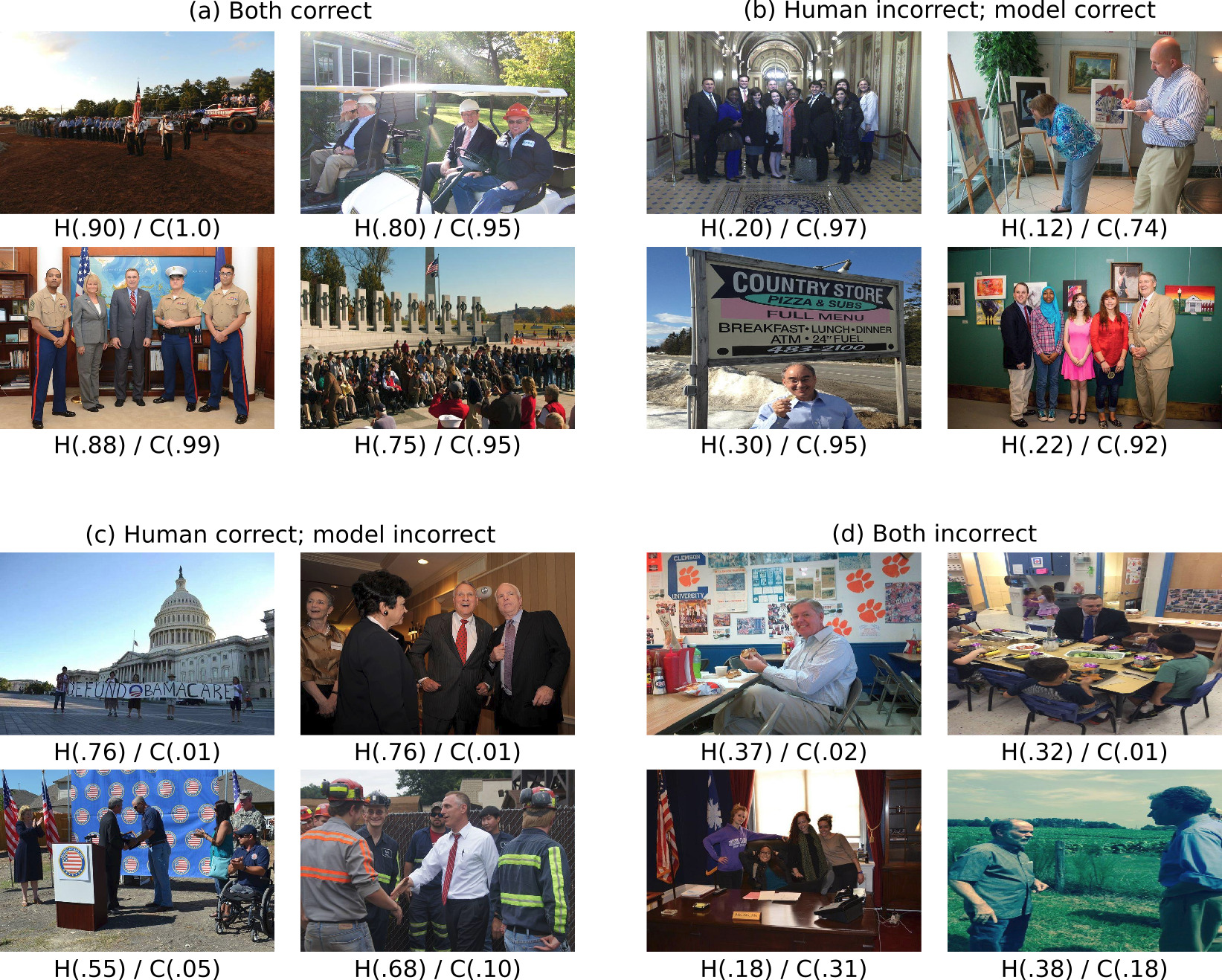}
    \caption{Comparison of party classifications by human judges and our model for \textbf{Republicans}. A higher score indicates Republican (H: Human, C: Classifier). (a) Both identified key features correctly (\eg, military officers, seniors). (b) Humans made wrong associations (\eg, non-white). (c) The model could not recognize text (\eg, ``DEFUND OBAMACARE'' or people. (d) Some Republicans exhibit features and activities similar to Democrats (\eg children). }
    \label{fig:rep-comp}
\end{figure*}

Figure~\ref{fig:dem-comp} shows the comparison of party classifications made by human judges and our trained model. Humans and the model correctly identified key features associated with Democrat politicians such as children, an LGBT symbol (rainbow), and minorities (Latino). 

The model identifies some ideological features humans do not.  For instance, the Capitol Building more frequently appears in Republican politicians' photo timelines, signaling authority, and the human judges may have not inferred the relation. In opposite cases, humans can use prior knowledge to infer true meanings of symbols that are infrequently used in the photographs and so cannot be automatically learned by our model from the given data. For example, a group of people wearing red t-shirts was recognized as Republicans by the model but they were volunteers from a nonprofit organization, Habitat for Humanity, judging from the logo on their shirts. In other cases, some Democrats' photographs depict activities which are typically associated with Republicans (\eg, military officers), and both human judges and the model failed to classify them correctly. Figure~\ref{fig:rep-comp} shows the comparison made on the Republican subset. Overall, humans rely more on \textbf{stereotyped gender and race} (\eg, `black' $\rightarrow$ Democrats) but are able to infer deeper and less frequent associations compared to the model prediction.

\begin{table}[]
\caption{Confusion Matrix for Party Classification by Human}
\begin{tabular}{l|l|c|c|c}
\multicolumn{2}{c}{}&\multicolumn{2}{c}{Ground Truth}&\\
\cline{3-4}
\multicolumn{2}{c|}{}&DEM & GOP &\multicolumn{1}{c}{Total}\\
\cline{2-4}
%\multirow{2}{*}{Human Classification}& DEM & $a$ & $b$ & $a+b$\\
Human & DEM & 72 (26.5\%) & 42 (15.4\%) & 114 (41.9\%)\\
\cline{2-4}
Guess & GOP & 64 (23.5\%) & 94 (34.6\%) & 158 (58.1\%)\\
\cline{2-4}
\multicolumn{1}{c}{} & \multicolumn{1}{c}{Total} & \multicolumn{1}{c}{136 (50\%)} & \multicolumn{    1}{c}{136 (50\%)} & \multicolumn{1}{c}{$272$}\\
\end{tabular}
\label{tab:human-comp}
\end{table}

\begin{table}[]
\caption{Confusion Matrix for Party Classification by Model}
\begin{tabular}{l|l|c|c|c}
\multicolumn{2}{c}{}&\multicolumn{2}{c}{Ground Truth}&\\
\cline{3-4}
\multicolumn{2}{c|}{}&DEM & GOP &\multicolumn{1}{c}{Total}\\
\cline{2-4}
%\multirow{2}{*}{Human Classification}& DEM & $a$ & $b$ & $a+b$\\
Model          & DEM & 88 (32.3\%) & 51 (18.8\%) & 139 (51.1\%)\\
\cline{2-4}
Output & GOP & 48 (17.6\%) & 85 (31.3\%) & 133 (48.9\%)\\
\cline{2-4}
\multicolumn{1}{c}{} & \multicolumn{1}{c}{Total} & \multicolumn{1}{c}{136 (50\%)} & \multicolumn{    1}{c}{136 (50\%)} & \multicolumn{1}{c}{$272$}\\
\end{tabular}
\label{tab:model-comp}
\end{table}

% \begin{table}[]
% \caption{Confusion Matrix for Party Classification by Model (on the whole set, $n=17,440$)}
% \begin{tabular}{l|l|c|c|c}
% \multicolumn{2}{c}{}&\multicolumn{2}{c}{Ground Truth}&\\
% \cline{3-4}
% \multicolumn{2}{c|}{}&DEM & GOP &\multicolumn{1}{c}{Total}\\
% \cline{2-4}
% %\multirow{2}{*}{Human Classification}& DEM & $a$ & $b$ & $a+b$\\
% Model          & DEM & 4,922 (28.2\%) & 3,374 (19.3\%) & 8296 (47.5\%)\\
% \cline{2-4}
% Output & GOP & 3,728 (21.4\%) & 5,416 (31.1\%) & 9144 (52.5\%)\\
% \cline{2-4}
% \multicolumn{1}{c}{} & \multicolumn{1}{c}{Total} & \multicolumn{1}{c}{8650 (49.6\%)} & \multicolumn{    1}{c}{8790 (50.4\%)} & \multicolumn{1}{c}{17440}\\
% \end{tabular}
% \label{tab:model-comp-whole}
% \end{table}

\section{Conclusion}
In this paper, we attempt to understand how members of Congress project political ideology through the images that they post on Facebook. For over a century, the systematic study of political ideology has been confined to political speeches, voting behavior  and written documents. While understanding how ideology is conveyed through these means is important, these media lack the emotional salience, persuasive power and compactness of images as means of political rhetoric. This research not only sheds light on how political ideology is expressed by members of Congress through visual means, but the methods developed here pave the way for entirely new avenues of study in social science. For instance, what types of ideological messages encoded in photographs shape public opinion about important political phenomena such as war or protest activity? How do politicians use images as a means of positioning themselves ideologically to their constituents during election cycles? Do ideological messages contained in images influence voter turnout and election results themselves? If so to what extent? Our work here represents a starting point that will allow scholars to answer these and other questions related to visual political communication and persuasion.

Validation using NOMINATE scores demonstrates that measures of the ideological content of images that we provide proxy well for political ideology measured via more traditional channels.
While the sample that we used for these analyses is by no means complete, our research advances the study of the visible aspects of political ideology and political rhetoric, presenting both a new method of estimating the ideological content of images along the left--right political spectrum and providing insights into the ideological content of images. 

In the past decade, computer vision and deep learning have enabled a wide range of applications in many domains. Our study further demonstrates its utility as a tool to diagnose subtle characteristics in human behaviors, especially related to political dimensions. As a huge volume of data describing human social and political activities, at the individual and societal level, are available from social media, automated and computational approaches will make more systematic and scalable analysis possible.

%\vspace{-10pt}

%\newpage
\section{Acknowledgement}
This work was supported by the National Science Foundation \#1831848, Hellman Fellowship, and UCLA Faculty Career Development Award. 

\bibliographystyle{aaai}
\bibliography{main}

% \end{document}

%\begin{acknowledgements}
%\end{acknowledgements}

% % BibTeX users please use one of
% \bibliographystyle{spmpsci}      % mathematics and physical sciences
% \bibliographystyle{spphys}       % APS-like style for physics
% \bibliography{main}   % name your BibTeX data base

% % Non-BibTeX users please use
% \begin{thebibliography}{}
% %
% % and use \bibitem to create references. Consult the Instructions
% % for authors for reference list style.
% % Format for Journal Reference
% Author, Article title, Journal, Volume, page numbers (year)
% Author, Book title, page numbers. Publisher, place (year)
% \end{thebibliography}

\newpage

\end{document}